\documentclass{article}

\usepackage{arxiv}

\usepackage[utf8]{inputenc} 
\usepackage[T1]{fontenc}    
\usepackage{hyperref}       
\usepackage{url}            
\usepackage{booktabs}       
\usepackage{amsfonts}       
\usepackage{nicefrac}       
\usepackage{microtype}      
\usepackage{lipsum}
\usepackage{graphicx}
\usepackage{natbib}
\usepackage{amssymb}

\usepackage[dvipsnames,svgnames,x11names]{xcolor}
\usepackage{algorithm}
\usepackage{algorithmicx}
\usepackage[noend]{algpseudocode}
\usepackage{listings}
\usepackage{amsmath}    

\graphicspath{ {./images/} }

\title{Gaussian process modelling of infectious diseases using the Greta software package and GPUs}

\date{}

\author{
Eva Gunn \\
  School of Mathematics and Statistics\\
  University of St Andrews\\
  St Andrews, KY16 9SS \\
   \And
 Nikhil Sengupta \\
  School of Computer Science\\
  University of St Andrews\\
  St Andrews, KY16 9SX \\
  \And
 Ben Swallow \\
  School of Mathematics and Statistics\\
  University of St Andrews\\
  St Andrews, KY16 9SS \\
  \texttt{bts3@st-andrews.ac.uk} \\
}

\begin{document}

\maketitle



\begin{abstract}
Gaussian process are a widely-used statistical tool for conducting non-parametric inference in applied sciences, with many computational packages available to fit to data and predict future observations. We study the use of the Greta software for Bayesian inference to apply Gaussian process regression to spatio-temporal data of infectious disease outbreaks and predict future spread. Greta builds on Tensorflow, making it comparatively easy to take advantage of the significant gain in speed offered by GPUs. In these complex spatio-temporal models, we show a reduction of up to 70\% in computational time relative to fitting the same models on CPUs. We show how the choice of covariance kernel impacts the ability to infer spread and extrapolate to unobserved spatial and temporal units. The inference pipeline is applied to weekly incidence data on tuberculosis in the East and West Midlands regions of England over a period of two years.
\end{abstract}




\section{Introduction}
\label{sec:intro}

Gaussian process regression is a non-parameteric stochastic statistical approach for modelling, inferring and predicting from data. Their flexibility enables them to be considered for a variety of tasks in scientific inference and prediction. One of their principal benefits is their inherent analytical tractability, allowing uncertainty in the process to be directly calculated from the calibration process.

Gaussian processes (henceforth GPs) have been used previously in infectious disease modelling, both as predictive data-driven models of spatial spread and incidence, as well as to form the basis of surrogate models or emulators of mechanistic models of infectious disease spread.

Whilst their analytical tractability facilitates fitting of GPs, inference and prediction requires complex linear algebra operations of often high-dimensional covariance matrices, making computation slow and cumbersome for many realistic data. A wide variety of approaches have been developed to improve efficiency and enable practitioners to approach larger scale problems common to infectious disease modelling {\cite[e.g.,][and references therein]{fastGPs,fifa}. Approaches to improve computation reduce the overhead of the problem by providing approximately independent subsets of the data through sparse approximations to the covariance kernel. The linear algebra operations can then be applied repeatedly on smaller subsets, a task that is well-suited to both parallelisation and fast computation on Graphical Processing Units (GPUs). Careful selection of these subsets can reduce unwarranted approximation errors in this.}

Previous authors have shown the wide variety of approaches that GPs can be used for in infectious disease modelling. For example, \citet{keelinggps} used GPs to estimate growth rates, $r$, across an area of England in the  SARS-CoV-2 pandemic, using a Mat\`{e}rn covariance kernel. {\cite{schistomaiasis} embed a GP as a latent process within a Poisson regression to interpolate snail distributions and allowing risks of schistomiasis to be estimated in Malawi.}

Other authors use GPs to emulate the behaviour of mechanistic models, such as SIR models and their variants. Based on careful runs at training points in the parameter space, the GP emulators can interpolate the behaviour of the mechanistic model at other to provide a computationally cheap approximation to the model surface. \citet{DUNNE2022} use GP emulators to calibrate the mean and variance of an age-structured compartmental model of asymptomatic spread of SARS-CoV-2 in hospitals, whilst \citet{grapKINGHAMJEFFERY2018111} and \citet{trostle24} directly model a moment-closure approximation to the stochastic jump process of an SIR model, with the latter extending to a spatial context. {Relatively little work has been done previously on the use of GPUs for infectious disease models, both due to the cost of the hardware making them prohibitively expensive to the average researcher, and also a steep learning curve in making general purpose software difficult to implement on them. There is also not a guarantee that they will provide significant improvement in all scenarios\cite{calgpu}. However, costs have reduced and they are now more readily available, with new software allowing integration of standard algorithms. Where these hardware have been used in infectious disease models, it has largely been in mechanistic individual based models where the GPU simulations enable repetition of computations across individuals giving significant speed improvements\cite[e.g.,][]{cudaibm,hybridgpuibm}.}

In section \ref{sec:methods}, we introduce the GP model structure for machine learning and the Bayesian workflow required for successful fitting, assessment and prediction using the \texttt{R} package \texttt{greta.gp} utilising tensorflow-probability on a GPU. In \ref{sec:casestudy}, we apply the methodological workflow to a case study of Tuberculosis incidence in the Midlands region of England. Finally in Section \ref{sec:concs}, the implications of the work and areas for further extensions are discussed.

\section{Gaussian processes}
\label{sec:methods}

A Gaussian Process is a collection of random variables, any finite number of which have (consistent) joint Gaussian distributions \cite{Rasmussen2004}. Gaussian processes are often used to model a smooth function $f$ of data $X$. In this case, the collection of random variables is the set $\{f(x): x \in X\}$, and is a Gaussian Process provided any finite subset, $\{f(x_1), \hdots, f(x_n)\}$ for some $x_1, ..., x_n \in X$, has a multivariate normal distribution with mean function $m(\cdot)$ and kernel function $k(\cdot,\cdot)$. That is: 

\begin{gather}
\label{eqn:GP}
 \begin{bmatrix} f(x_1) \\ \vdots \\f(x_n) \end{bmatrix}
 \sim
 N\Biggl(
  \begin{bmatrix}
   m(x_1) \\
   \vdots \\
   m(x_n)
   \end{bmatrix},
   \begin{bmatrix}
   k(x_1, x_1) & \hdots & k(x_1, x_n) \\
   \vdots & \hdots & \vdots\\
   k(x_n, x_1) & \hdots & k(x_n, x_n)
   \end{bmatrix}
   \Biggr)
\end{gather}

\noindent Given \ref{eqn:GP}, we can write

\begin{equation}
    f \sim \text{GP}(m(\cdot), k(\cdot,\cdot)). \nonumber
\end{equation}

\subsection{Mean \& kernel structures} \label{sec:mean_ker_structures}

A GP is fully determined by its mean and kernel function. The mean function, $m(\cdot)$, is the expected value of the function. It is often taken to be $\mathbf{0}$ for simplicity, and the data can be normalised for this to work. Alternatively, a constant mean function (or bias term) can be estimated from the data, centring the process away from zero.


The kernel function, $k(\cdot,\cdot)$, describes the covariance structure between pairs of data points. It can be any positive definite function, however, typically standard kernel functions are used because of their convenient properties. The kernels tend to have two parameters: the variance $\sigma^2$ and length scale $l$ parameters. The variance parameter indicates how much the function deviates from the mean, and the length scale parameter indicates how quickly the correlation between points decreases with distance. As the length scale increases the function becomes smoother and points further apart are relatively more correlated. 

Here we detail four types of kernel functions, namely Exponential, Mat\`{e}rn, Radial Basis Function and Periodic kernels. The first three are characterised by their varying smoothness and the last is used for functions that exhibit periodic behaviour. Suppose that the vectors $\mathbf{x}$, $\mathbf{x'}$ represent two different (potentially-multivariate) data points and let $\lVert \mathbf{x-x'} \rVert$ represent the Euclidean distance between them, then the following kernels are defined as:

\begin{enumerate}
\item[] \emph{Exponential Kernel}

\begin{equation}
k_{exp}(\mathbf{x},\mathbf{x'}) =\sigma_{exp}^2\text{exp}(-\frac{\lVert \mathbf{x-x'} \rVert }{2l_{exp}}) 
\label{eq:exponential_kernel} \nonumber
\end{equation}

\item[] \emph{Mat\`{e}rn kernel}

The Mat\`{e}rn kernel function takes an extra parameter $\nu$, so generates a class of kernels rather than a single kernel. $\nu$ must be positive and the function is simple in the cases $\nu = p + 1/2$ with p a positive integer. We only consider the case $\nu = 3/2$.

\begin{equation}
k_{\nu = 3/2}(\mathbf{x},\mathbf{x'}) = \sigma_{\nu = 3/2}^2(1+\frac{\sqrt{3}\lVert \mathbf{x-x'} \rVert}{l_{mat}})\text{exp}(-\frac{\sqrt{3}\lVert \mathbf{x-x'} \rVert}{l_{mat}})\nonumber
\end{equation}

\item[] \emph{Radial Basis Function (RBF) kernel for vector}

\begin{equation}
k_{rbf}(\mathbf{x},\mathbf{x'}) = \sigma_{rbf}^2\text{exp}(-\frac{1}{2l_{rbf}^2}\lVert \mathbf{x-x'} \rVert^2 )\nonumber
\end{equation}

\item[] \emph{Periodic kernel}

\begin{equation}
    k_{per}(\mathbf{x},\mathbf{x'}) = \sigma_{per}^2\text{exp}(-\frac{\text{sin}^2(\pi\lVert \mathbf{x-x'} \rVert/p)}{2l_{per}^2})\nonumber
\end{equation}

\end{enumerate}

\noindent Note that slight variations of these kernel definitions exist, but the specifications above are those used in the software package that we use in our analysis. 

\begin{figure}[h]
    \centering
    \includegraphics[width=1\textwidth]{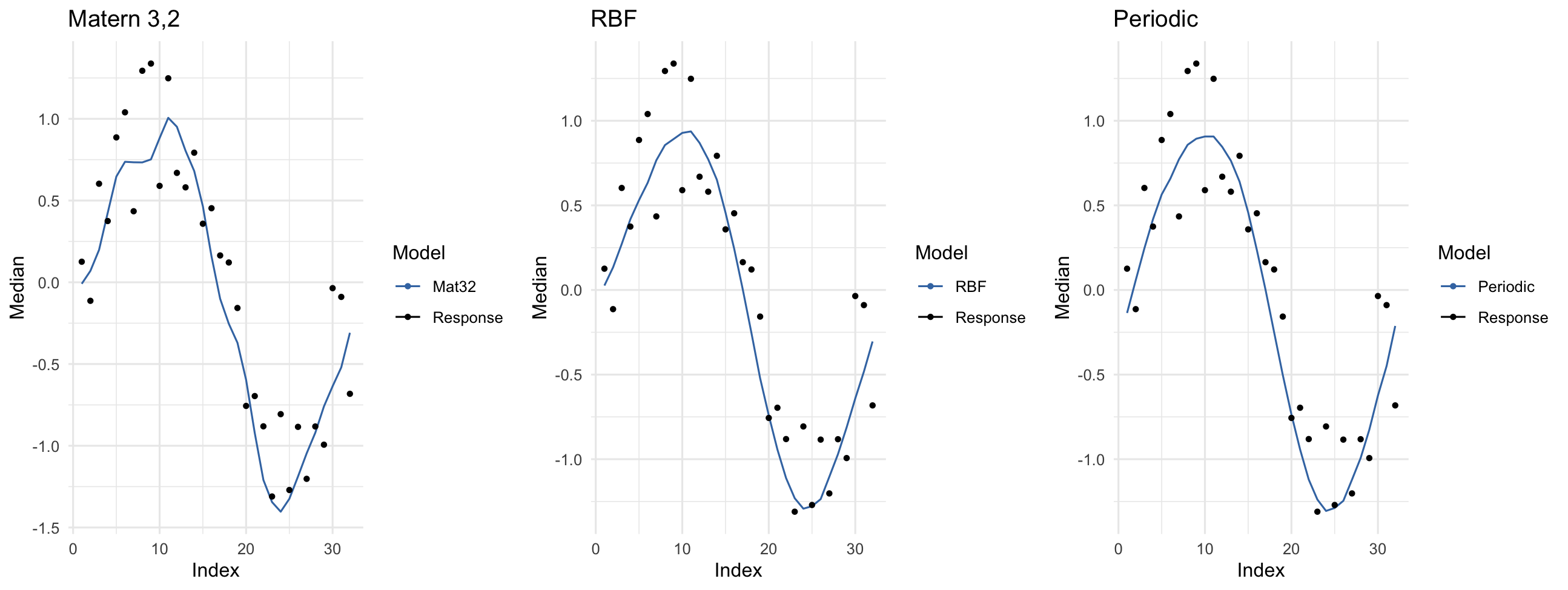}
    \caption{Comparison of smoothness of varying kernels (Mat\`{e}rn, radial basis function and periodic) for some synthetic data.}
    \label{fig:kernelcomp}
\end{figure}

It can be shown that the first three functions are related. The Mat\`{e}rn kernel function is a generalisation of the exponential and radial basis functions. When $\nu = 1/2$, the kernal corresponds to a variation of the exponential function defined here and as $\nu \rightarrow \infty$, the kernel approaches the radial basis function. The parameter $\nu$ controls the smoothness of the function. When $\nu$ is low, such as in the case of the exponential function, the function is rough. As $\nu$ increases the function becomes smoother. In the case of the radial basis function it is actually infinitely differentiable. It is not obvious what level of smoothness to select prior to trialling the different kernel functions. In this paper, we experiment to see how choosing different smoothness of kernels affected goodness of fit to the data and ability to make predictions.

A useful property of kernel functions is that both the sum and product of two kernels is also a kernel \citep{Rasmussen2004}. This makes modelling spatio-temporal processes easier as, assuming separability, we can assign different kernel functions to the space and time components of the model. Additive kernel functions are generally used for modelling the independent effects of the explanatory variables, whilst  multiplicative kernels are generally used for the interactive effect between variables.

\subsection{Combined kernel functions}

 In line with section \ref{sec:mean_ker_structures}, we add together a kernel function for the time component ($k_{time}$); a kernel function for the space component ($k_{space}$); and a multiplicative kernel for the interaction between space and time ($k_{space-time} = k_{time}* k_{space}$) to get a kernel function that is spatially and temporally indexed:

 \begin{equation}
     k = k_{time} + k_{space} + k_{space-time}\nonumber
 \end{equation}
 
\noindent This type of approach has been used in previous research, as by \cite[e.g.,][]{Senanayake_OCallaghan_Ramos_2016,albinati17,hawryluk21}.

\subsection{Observation process}

{In many situations, the Gaussian process is capable of capturing the underlying spatio-temporal latent structure in the data, however is inappropriate to directly model the observed data. For example, disease incidence is often measured through counts of new infections, a discrete random variable. Therefore, it is necessary to embed the Gaussian process within a probability mass (discrete) or density (continuous) function to map from a continuous valued output to an appropriate support, as well as account for additional errors or uncertainties in the data collection.}

In larger populations, we would generally expect a higher absolute number of infections. As a result, it is common to split the mean number of infections ($\mu_{ij}$) as a product of the background population effect ($e_{ij}$) and the excess risk ($\theta_{ij}$) \citep{LAWSON_2023} as to focus on modelling the excess risk:

\begin{equation}
	\mu_{ij} = e_{ij}\theta_{ij}. \label{eq:expectedinfec}
\end{equation}

The background population effect ($e_{ij}$) is the number of cases you would expect for a given population size. It is is the product of the population size ($p_{ij}$) and the crude incidence rate (R), which is the number of new infections per person in the whole population over the time period.

\begin{equation}
e_{ij} = p_{ij}R\nonumber
\end{equation}

\begin{equation}
R = \frac{\sum_i \sum_j y_{ij}}{\sum_i \sum_j p_{ij}}\nonumber
\end{equation}

\noindent For the excess risk, we take the logarithm of the excess risk, since $\theta_{ij} > 0$, and model it using a Gaussian Process:

\begin{equation}
\log(\theta_{ij})  = f(x_{ij})\nonumber
\end{equation}
\begin{equation}
f(.) \sim GP(m(.), k(.,.)).\nonumber
\end{equation}

{
\noindent The relationship between the data and the Gaussian process is hence:

\begin{equation}
	y_{ij} \sim \text{Pdf}(\mu_{ij}, \phi_{ij}),\label{eq:obsmodel}
\end{equation}

\noindent where $Pdf(.)$ is the selected observation response distribution with mean $\mu_{ij}$, as defined as in equation \ref{eq:expectedinfec}, and $\phi_{ij}$ are additional distribution-specific parameters, such as variance or dispersion parameters.}

\subsection{Model fitting}
\subsubsection{Inferring parameters}

{Once the model structure in equation (\ref{eq:obsmodel}) has been defined,} the parameters of the response distribution ($\phi_{i,j}$), as well as the GP mean and kernel functions need to be estimated from the data. The most common ways are maximum likelihood estimation \citep[e.g.,][]{mardia1984maximum} and Bayesian inference \citep[e.g.,][]{mackay1992bayesian}{, the latter requiring additional specification of prior distributions for the model parameters}. Using a Bayesian Markov chain Monte Carlo framework \citep[see][for more details]{mcmcandrieu} is more computationally intensive because of the necessity to repeatedly invert the covariance matrix to obtain the posterior distribution, however it provides a straightforward means of assessing parameter uncertainty through the posterior distribution. Understanding uncertainty is particularly important in an epidemic context to enable associated risks to be taken into consideration. {This uncertainty naturally reduces in areas where data are particularly informative, whilst allowing forecasts to be less informed in areas of space and/or time of data sparsity.}

\subsubsection{Model checking and comparison}


In order to assess fitted models alignment to the data and within and out of sample prediction, posterior predictive checks can be conducted. These approaches compare the density of the predictive and observed values, and secondly by applying a discrepancy measure {on observed and model-fitted data or observed and data simulated from the fitted model} to see if they give similar values. In the latter case we use the Freeman-Tukey statistic (\ref{eq:Tukey}), which is less sensitive to smaller values than for example the Chi-Squared statistic \citep{conn2018}. If the model is a good fit then a plot of the observed and simulated Freeman-Tukey values should show even spread over the $y = x$ line. The Freeman-Tukey statistic is calculated as:

\begin{equation}
T(\pmb{y},\pmb{\theta})=\sum_{i}(\sqrt{y_{i}}-\sqrt{E(y_{i}|\pmb{\theta}})
)^2,\nonumber
\label{eq:Tukey}
\end{equation}

\noindent{where $y_i$ are the observed data and $E(y_{i}|\pmb{\theta})$ are the fitted model values for those data.}

To compare {between models with differing response distributions or kernel functions}, three metrics designed for Bayesian inference will be utilised: leave-out-one cross validation, continuous ranked probability score and a Bayesian p-value. They each serve different purposes in model comparison. Our first metric was an approximate leave-out-one cross validation (LOO-CV) {\cite{vehtariloo}. This approach aims to fit the model to a subset of the data and use this fitted model to predict the removed observations.} LOO-CV conducted over all posterior samples and data points would be very time consuming, so approximations can be made, which use the log-likelihood of the data given posterior draws of the distribution's parameters. Such an approximation is used by the loo package in \texttt{R}, and gives an information criteria `looic' and is described in Algorithm \ref{alg:loo}. As with other information criteria, the lower the looic score the better the model fits the data. looic records how well the model performs when making predictions on the training data.

\begin{algorithm}
    \caption{Loo Algorithm} \label{alg:loo}
    \begin{algorithmic}[1]
        \State Make S = 200 posterior simulations of f, $\phi$ and $\lambda$
        \State Make an empty matrix of dimension S by N (\# data points = 6760)
        \For{$i$ in 1:S}
            \For{$j$ in 1:N}
                \State Calculate $\mu$ using equation \ref{eq:expectedinfec}
                \State Convert $\mu$ and $\phi$ to $r$ and $p$ using equations \ref{eq:r} and \ref{eq:p}
                \State Calculate $\pi$ using equation \ref{eq:pi}
                \State Calculate log-likelihood of observed values given $\pi$, $r$ and $p$
                \State Store to position i, j in the matrix
            \EndFor
        \EndFor
        \State Calculate relative effective sample sizes from the log-likelihood matrix using relative\_eff from loo package.
        \State Calculate loo using the relative effective sample sizes and log-likelihood matrix using loo from loo package.
    \end{algorithmic}
\end{algorithm}

The second metric, continuous ranked probability score (CRPS), considers how well the model can predict new data. If the model is a good fit, one would expect the CDF of the predictions to be similar to a function that is 1 if the prediction is greater than or equal to the observed value and 0 otherwise \cite{Faran_2023}. If the model is good at making predictions then the difference in area between these two functions should be close to 0.


\noindent Given this, let y be the observed value, x the predicted variable, and so F(x) its CDF function, then the CRPS is defined as:

\begin{equation}
crps(F,y) = \int_\mathbb{R}[F(x)-\mathbb{1}(x \geq y)]^2dx\nonumber
\end{equation}

\noindent This was shown by \cite{doi:10.1198/016214506000001437} to be equivalent under certain conditions (finite first moment) to:

\begin{equation}
crps(F,y) = E_X |X - y|-\frac{1}{2}E_{X,X'}|X-X'|\nonumber
\end{equation}

\noindent where $X, X'$ are draws of the observation from the posterior distribution such that $X$ and $X'$ are independent and from the same distribution.  This is the form that is used by the {\texttt{loo}} package in \texttt{R}. It is averaged across the samples to give a single metric. 

{\begin{algorithm}
    \caption{CRPS Algorithm} 
    \label{alg:crps}
    \begin{algorithmic}[1]
        \State Make 1000 predictions for each location for each week 105 to 108
        \For{week in 105:108}
            \State $y\_pred \gets$ predictions[,time $==$ week]
            \State $X \gets$ $y\_pred$[1:500,] \Comment{split into two predictive samples}
            \State $X' \gets$ $y\_pred$[,501:1000] 
            \State Calculate CRPS using X, X' and y (the observations) via the crps function from the loo package.
        \EndFor
    \end{algorithmic}
\end{algorithm}}

The third metric is the Bayesian P-value, {Algorithm \ref{alg:bayesp}, aims to determine whether predictive data simulated from the fitted model shows similar characteristics to the osberved data}. It is a formalisation of the posterior predictive check using a specified discrepancy measure. From \cite{gelman2013bayesian} the Bayesian p-value is defined as:

\begin{equation}
    p_B = Pr(T(y^{rep}, \boldsymbol{\theta}) \geq T(y,\boldsymbol{\theta})|y)\nonumber,
\end{equation}

\noindent  where T is the test statistic and the probability is taken over the posterior distribution of $\boldsymbol{\theta}$ and the posterior predictive
distribution of $y$ ($y_{rep}$). A good p-value is one that is close to 0.5; although anywhere between 0.05 and 0.95 is deemed acceptable \citep{gelman2013bayesian}. Here we again use the Freeman-Tukey test statistic as $T$. 

{\begin{algorithm}
    \caption{Bayesian P-value} 
    \label{alg:bayesp}
    \begin{algorithmic}[1]
        \State Tukey.obs $\gets$ Freeman-Tukey Statistic (equation \ref{eq:Tukey}) in terms of the observed and expected value
        \State Tukey.sim $\gets$ Freeman-Tukey Statistic in terms of the simulated response and expected value
        \State Make 1000 simulations of Tukey.obs and Tukey.sim using posterior parameter draws
        \State Bayes\_p $\gets$ mean(Tukey.obs$>$Tukey.sim)
    \end{algorithmic}
\end{algorithm}}

\subsection{Prediction}

Gaussian Processes make generating predictions relatively simple since they have a tractable form for the posterior predictive distribution. If the observed data is the matrix $\mathbf{X}$ and the predicted new data is $\mathbf{X_*}$, such that  $\mathbf{f}$ and $\mathbf{f}_*$ is the function on the old and new points respectively, then the joint distribution is \citep{Rasmussen2004}:

\begin{equation}
    \begin{bmatrix}
        \mathbf{f}\\
        \mathbf{f}_*
    \end{bmatrix} 
    \sim N \Biggr(
    \begin{bmatrix}
        m(\mathbf{X}) \\
        m(\mathbf{X}_*)
    \end{bmatrix}
    \begin{bmatrix}
    \mathbf{K} & \mathbf{K}_* \\
    \mathbf{K}_*^T & \mathbf{K}_{**}
    \end{bmatrix}
    \Biggl)\nonumber
\end{equation}

\noindent with $\mathbf{K} = k(\mathbf{X}, \mathbf{X})$, $\mathbf{K}_* = k(\mathbf{X}, \mathbf{X_*})$, $\mathbf{K}_{**} = k(\mathbf{X}_*, \mathbf{X}_*)$. Predictions are then made by obtaining the marginal distribution of $f$ from the joint distribution. This is calculated as follows:
\begin{equation}
    \mathbf{f}_*|\mathbf{f},\mathbf{X},\mathbf{X}_* \sim N(m(\mathbf{X}_*) + \mathbf{K}_*^T\mathbf{K}^{-1}(\mathbf{f}-m(\mathbf{X})), \quad \mathbf{K}_{**} - \mathbf{K}_*^T\mathbf{K}^{-1}\mathbf{K}_*).\nonumber
\end{equation}

\subsection{Approximations}

The use of Gaussian Processes in practice can be limited by the computational complexity of calculating the inverse for the kernel, which increases with $O(n^3)$ as $n$ (the number of training cases) increases \citep{JMLR:v6:quinonero-candela05a}.  The inversion is necessary to make predictions and calibrate parameters. Hence, when the data is large (more than $\sim$ 2500) \citep{Senanayake_OCallaghan_Ramos_2016} it is often best to take a subset of $m < n$ data points, which is known as a sparse approximation. How this subset is used depends on the approximation method.

There are several methods of sparse approximation. The baseline approximation is the Subset of Data (SoD) Approximation. This method simply takes a subset $m$ of $n$ of the training data. This reduces the computational complexity to $O(m^3)$, but ignores a lot of the data. The method that \texttt{greta.gp} has, and that used in this analysis, is the Subset of Regressors (SoR) Approximation. Subset of regressors was proposed by \cite{wahba1990spline, poggio1990networks} and adapted by \cite{smola2001sparse}. It is a reduced rank method, that is it approximates the original kernel functions by a weighted sum of the  kernel functions applied to the subset. Let the original n functions be written 
$S_\mathcal{N} = \{k(x,x_i|\theta) : i \in \mathcal{N}\}$ with $\mathcal{N} = \{1,2,..., n\}$. If we select the inducing variables (indices)
$\mathcal{M} \subset \mathcal{N}$  such that $S_\mathcal{M} = \{k(x,x_j|\theta) : j \in \mathcal{M}\} \subset S_\mathcal{N}$, then we write the approximation as $\hat{k}(x,x_r|\theta) = \sum_{j\in \mathcal{M}}c_{jr}k(x,x_j)$ 
for $r \in \mathcal{N}$. {The coefficients are found by solving the equation $k(x_j,x_j)^{1/2\intercal}\mathbf{c}=S_\mathcal{M}$. A disadvantage of this approximation, is that it may underestimate the predictive variance when a data point is far away from points in the subset \citep{JMLR:v6:quinonero-candela05a}. See \citep{JMLR:v6:quinonero-candela05a} for details on other common approximations.}

The subset of the inputs used (inducing inputs) need to be chosen carefully as these change the final solution (while the inducing variables do not) \citep{JMLR:v6:quinonero-candela05a}. Three common ways of choosing inducing inputs are greedy selection, grid selection, and K-means ++ \citep{GalyFajou2021AdaptiveIP}. Greedy selection, suggested by \cite{titsiasgreedy}, adds points iteratively if they give the best fit in a batch. This method is extremely computationally expensive, as it involves fitting the Gaussian Process many times. K-Means ++ is good at taking into account the structure of the data, but is also computationally intensive.

\subsubsection{Use of Greta software in R}

The GPs are fitted in \texttt{R} \citep{R-Core-Team:2021aa} using the package \texttt{greta} \citep{greta}, which was designed with three main improvements in mind over existing software \citep{Golding2019}: analysis can be carried out directly in \texttt{R}; easy to extend the package in \texttt{R}; and fast inference for large datasets using TensorFlow \citep{tensorflow2015-whitepaper}. In this paper, we also used an extension of the \texttt{greta} package, \texttt{greta.gp} \citep{greta.gp}, which facilitates the fitting of GPs. {Additions to the package were required, to both enable the zero-inflated probability mass functions appropriate for the data, as well as extensions to some of the kernel functions to provide the required flexibility. The periodic kernel did not allow you to choose which covariate to act on as it had only previously been used on time series data. The Tensorflow code was extended the code so that it can be applied to spatio-temporal data. These changes can be accessed through the forked versions of greta and greta.gp on the author's GitHub. See the Supporting Information for instructions on how to make your own changes to greta and greta.gp}.

There are currently three sampling algorithms available in \texttt{greta}:  Random walk Metropolis-Hastings, Hamiltonian Monte Carlo, and a multivariate slice sampling algorithm. Random Walk can struggle with high dimension probability distributions, which is inherent in Gaussian Processes \citep{Titsias_mcmcgp11}. Although, slice sampling has been used for Gaussian Processes previously by \cite{murray2010slice}, we found that it sampled poorly when we trialled it. So we chose to use Hamiltonian Monte Carlo for all our models. This method requires you to select a minimum and maximum number of leapfrog steps. If the number of leapfrogs is low, samples will have high autocorrelation, but if high, it takes much longer to sample \citep{Arakawa805499}. We found for our models setting the minimum number to 15 and the maximum number 20 was a good trade-off between a slightly slower model and sampling well.

\section{Case study: Tuberculosis in the East and West Midland regions}
\label{sec:casestudy}

\subsection{Data}

The UK government releases weekly reports of notifiable infectious diseases (NOIDS) \citep{noids_data} in England and Wales. Several diseases are reported: Measles, Mumps, Rubella, Scarlet Fever, Whooping Cough, Tuberculosis, Acute Meningitis, Malaria, and food poisoning. The reports contain how many cases there were of each disease within Public Health England (PHE) Region; county; and local and unitary authorities. National park local authorities and some corporations in London were not included in the weekly reports. We chose to focus on Tuberculosis counts in the local and unitary authorities in the {65} East and West Midland PHE regions between 2022 and 2024, which had a significant number of counts to model and could demonstrate how Gaussian Process models can be used on current outbreaks. {This gave a total of 6760 training data points.} Note that the cases only include active Tuberculosis cases and not latent Tuberculosis cases. For training the Gaussian Process model we used the years 2022 and 2023, and for testing the predictive ability used the first month of 2024. This means we had 104 weeks of training data and 4 weeks of testing data. We also took data from the government on the longitude and latitude (and boundaries) \citep{lpa_data} and the population size \citep{ONS2022} of each local planning authority. 

\subsection{Likelihood} \label{sec:modelspec}

The data consists of Tuberculosis counts in authority $i$ {for $i \in {1, \hdots, 65}$} in week $j$ for $j \in {1, \hdots, 104}$ in the East and West Midlands, England. We model these counts using the Negative Binomial distribution, which has the following likelihood:

\begin{equation}
    f(x) = {x+r-1 \choose x}p^r(1-p)^x, \nonumber
\end{equation}

\noindent where $r$ is the number of successes and $p$ is the probability of success. In the context of modelling diseases it is often preferred to reparameterise in terms of the mean ($\mu$) and dispersion parameter ($\phi$) using the following:

\begin{equation}
    r = \frac{1}{\phi} \label{eq:r}\nonumber
\end{equation}

\begin{equation}
    p = \frac{r}{\mu + r}.  \label{eq:p}\nonumber
\end{equation}

\noindent The number of infections for each authority i in week j with Negative Binomial distribution is then:

\begin{equation}
	y_{ij} \sim \text{NB}(\mu_{ij}, \phi_{ij}).\nonumber
\end{equation}

Often, count data is also modelled using a Poisson distribution, which is in fact the limiting case of the Negative Binomial when $\phi \rightarrow \infty$. A small value of $\phi$ justifies the use of the more flexible distribution. 

\subsubsection{Zero-inflated likelihood}

Often, more zeroes occur in the data than would be expected for a Negative Binomial distribution. One way to account for this is by using a zero-inflated model, which has probability density function:

\begin{equation}
    f(x) =
    \begin{cases}
        \pi + (1-\pi)p^r & x = 0\\    
        (1-\pi){x+r-1 \choose x}p^r(1-p)^x & x > 0,
    \end{cases}  \nonumber
\end{equation} 

\noindent where $\pi$ is the probability of extra zeroes, p is probability of success and r is the number of successes. We trial this probability density function in addition to the standard Negative Binomial. As before, this can be reaparemeterised in terms of mean and dispersion parameter such that the number of infections for each authority i in week j is now modelled as:

\begin{equation}
	y_{ij} \sim \text{ZINB}(\pi, \mu_{ij}, \phi_{ij}).\nonumber
\end{equation}

\subsection{Priors}

Part of the Bayesian framework is setting prior distribution  on the model parameters. For the kernel parameters we use the priors recommended by the Stan handbook \citep{Stan_user_guide}, such that:

\begin{equation}
    l_K \sim \text{InvGamma}(5,5)\nonumber
\end{equation}
\begin{equation}
  \sigma_K \sim \text{N}(0,1)
      \label{eq:var_par}\nonumber
\end{equation}

\noindent for each kernel K with these parameters. For the Negative Binomial we chose the following:  

\begin{equation}
    \frac{1}{\sqrt\phi}\sim \text{N}(0,1)\nonumber
\end{equation}

\noindent also suggested by Stan in their Developer Wiki \citep{Stan-Dev_2023} which makes $\phi$ relatively small and consequently we have an overdispersed model. We also gave the mean function parameter the prior in (\ref{eq:var_par}).

For the zero-inflated models we have an extra parameter $\pi$ that gives the probability of extra zeroes. For this, we used the recommendation of \cite{inla_zi}, which suggested defining $\pi$ as a function of $\lambda$

\begin{equation} 
    \pi = \frac{\text{exp}(\lambda)}{1 + \text{exp}(\lambda)} \label{eq:pi}\nonumber
\end{equation}

\noindent with the following prior on $\lambda$:

\begin{equation}
    \lambda \sim \text{N}(-1, 5)\nonumber
\end{equation}

\subsection{Models} \label{sec:models}

Rather than building every combination of models from the set of kernel functions, {models are altered sequentially based on the best previous model. The focus is initially on optimising the temporal kernel through varying smoothness and also allowing a periodic kernel function, while keeping the spatial kernel fixed (Models 1-3). The Mat\`{e}rn(3,2) kernel is one with a mid-level smoothness as a starting point and often the standard kernel used in spatial statistics. Once the optimal temporal structure of those testes is found, namely a Mat\`{e}rn(3,2) added to a periodic kernel, then different spatial kernels are also tested (Models 4\&6).} The smoothness of the spatial kernel, {based on Euclidian distance between the latitude and longitude of each authority,} can be varied to further optimise model fit. Finally we test how changing the distribution to Zero-Inflated affects the model (Model 5). The exact choices of kernel functions can be seen in Table \ref{tab:kernel_structure}.

\begin{table}[h]
\caption{Kernel functions used}
  \centering
  \begin{tabular}{|c|c|c|c|c|}
    \hline
    \textbf{Model} & \textbf{$k_{time}$} & \textbf{$k_{space}$} & \textbf{$k_{space-time}$} & Distribution \\
    \hline
    1 & rbf & $\text{mat}_{32}$ & $k_{time}* k_{space}$ & Negative Binomial \\
    2 & $\text{mat}_{32}$ & $\text{mat}_{32}$ & $k_{time}* k_{space}$ & Negative Binomial\\
    3 & $\text{mat}_{32}$ + periodic & $\text{mat}_{32}$ & $k_{time}* k_{space}$ & Negative Binomial\\
    4 &  $\text{mat}_{32}$ + periodic  & rbf & $k_{time}* k_{space}$ & Negative Binomial \\
    5 & $\text{mat}_{32}$ + periodic & $\text{mat}_{32}$ & $k_{time}* k_{space}$ & Zero-Inflated Negative Binomial\\
    6 & periodic & $\text{mat}_{32}$ & $k_{time}* k_{space}$ & Negative Binomial\\
    \hline
  \end{tabular}
  \label{tab:kernel_structure}
\end{table}

A grid approach was used to defining the inducing points, because it is computationally cheap and most of the variation in our explanatory variables comes from the time component as the locations are the same throughout the data. We kept all the locations and sampled the time component every 5 weeks, with the final week also included as it is the most important week for prediction. So in total we used 22 time points, giving 1430 (65 locations * 22 weeks) inducing inputs. {The HMC algorithm was run for 2000 iterations, with the first 1000 discarded as burn-in. Since we used a Bayesian method we needed to check convergence of the chains to a stable distribution, running four independent chains. Brooks-Gelman-Rubin statistics were monitored until less than 1.1, supporting convergence \citep{Stephen1998} (see Supporting Information for example traceplots and the convergence statistics).}


\subsection{GPU setup}

Both Hamiltonian Monte Carlo sampling and Gaussian Process prediction involves repeated matrix operations, including multiplication and inversions, which are computationally intensive. A GPU can signifcantly speed this process up by parallelising these processes. In our results, we demonstrate the improvement in time taken using a GPU compared to a CPU.

Our experiments were conducted using a NVIDIA A30 GPU with 24GB HBM2 memory, 3,804 CUDA cores, and 224 tensor cores (up to 10.3 TFLOPS FP32). We also used a 13th Gen Intel(R) Core(TM) i7-13700K CPU with 16 cores, 24 threads, and a maximum clock speed of 5.4 GHz to compare the speed of running the models on the GPU to the CPU. To ensure compatability of the \textit{greta} package with the GPU, we used Python 3.10, CUDA toolkit 11.8.0, cuDNN version 8.9.7.28 and Tensorflow 2.13.0.

\subsection{Results}
\label{sec:results}

The six models from Table \ref{tab:kernel_structure} were fitted until converegence. There were two main arguments, warm-up samples and number of chains, that we tuned to get convergence for each model. More complex models often require more warm-up samples for convergence. In general, increasing the number of chains can reduce the time for convergence. However, increasing the number of chains increases the memory demand at one given time, and so more complex models may only allow for fewer chains. For each model, 1000 posterior samples were used post-convergence. Convergence was assessed through visual checking of trace plots and the Brooks-Gelman-Rubin statistic. Posterior statistics for Model 2 are summarised in Table \ref{tab:m2_param_ests}. 

The exponential kernel was initially trialled, that is one that even less smooth than the Mat\`{e}rn (3/2) kernel, however, the samples failed to mix well despite different starting positions and many samples. Hence, this was deemed unsuitable for our data. 

Comparisons were made in computational time on both CPU and GPU for both the simplest and most complex model, with around 60-70\% reduction in overall fitting time when using the GPU combined with Tensorflow (Table \ref{tab:timetaken}).

\begin{table}[h]
\centering
\begin{tabular}{|c|c|c|c|}
\hline
\textbf{Parameter} & \textbf{Posterior Median } & \textbf{Lower CI 95\%} & \textbf{95\% Upper CI }\\
\hline
len\_space   & 0.306    &    0.222   &     0.416 \\ \hline
sigma\_space & 1.048    &    0.788   &     1.398 \\ \hline
sigma\_time  & 0.234    &    0.069   &     0.667 \\ \hline
len\_time    & 1.096    &    0.439   &     3.823 \\ \hline
bias\_var    & 0.799    &    0.113   &     2.316 \\ \hline
phi         & 0.244    &    0.135   &     0.387 \\ \hline
\end{tabular}
\caption{Posterior summary statistics for Model 2 based on 1000 posterior samples.}
\label{tab:m2_param_ests}
\end{table}

\begin{table}[h]
\caption{Time taken for two models on CPU and GPU (rounded to nearest hour)}
  \centering
  \begin{tabular}{|c|c|c|}
    \hline
    \textbf{Model} & \textbf{Time taken on CPU} & \textbf{Time taken on GPU} \\
    \hline
    1 &  191 hours & 75 hours \\
    5 &  195 hours & 64 hours \\
    \hline
  \end{tabular}
  \label{tab:timetaken}
\end{table}

\begin{figure}[h]
    \centering
    \includegraphics[width=1\textwidth]{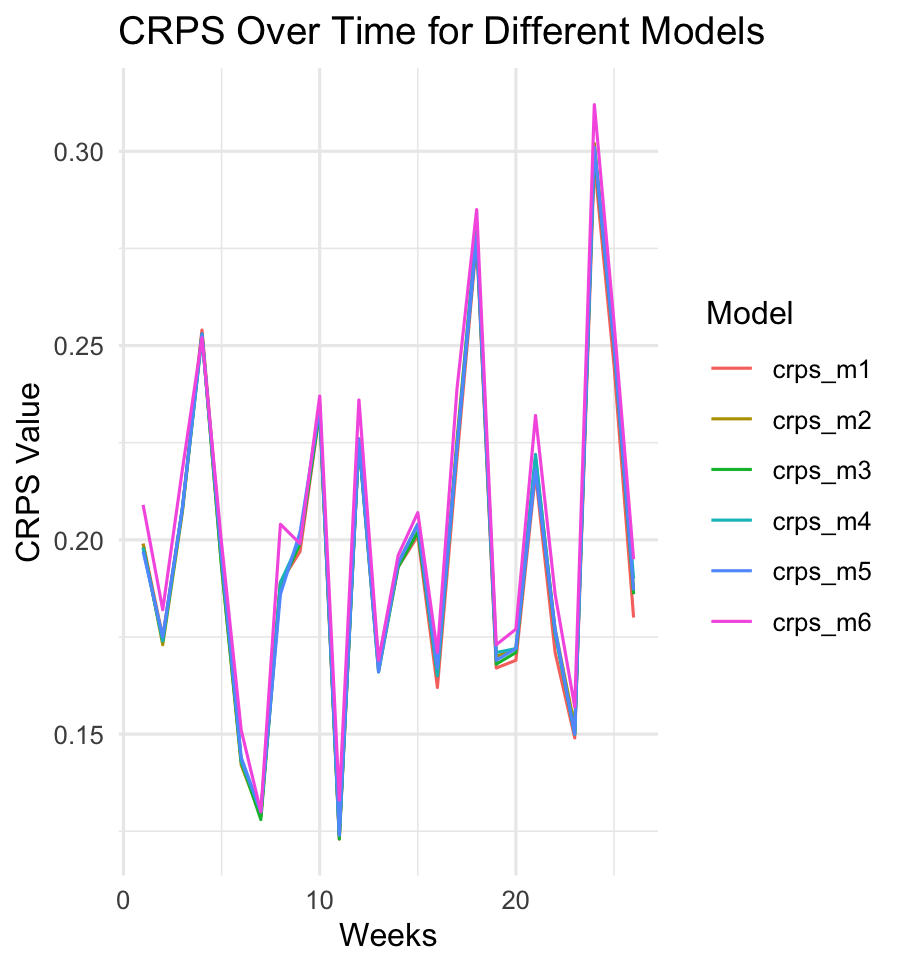}
    \caption{CRPS over weeks used to train models.}
    \label{fig:crps_long}
\end{figure}

Model comparison and assessment {was initially conducted using the LOOIC, Tukey statistic and CPRS over 4 weeks of prediction (see Supporting Information), but was extended to up to 26 weeks ahead (Table \ref{tab:loo_ic})}. CPRS across weeks is shown in Figure \ref{fig:crps_long}. There was relatively small differences across models, with Model 2 overall giving the best trade-off between model complexity and fit. Full summaries of the six models are presented in the Supplementary materials.

\begin{table}[h]
\centering
\begin{tabular}{|c|c|c|c|}
\hline
\textbf{Model} & \textbf{LOOIC} & \textbf{Tukey} &
\textbf{CRPS}\\ \hline
Model 1   &  4770.6 & 0.377 & 0.1957\\ \hline
Model 2   &  4768.8 & 0.612 & 0.1976 \\ \hline
Model 3   &  4771.6 & 0.426& 0.1967 \\ \hline
Model 4   &  4769.1 & 0.417& 0.1979 \\ \hline
Model 5   & 4773.0 & 0.314& 0.1973 \\ \hline
Model 6   & 4778.6  & 0.371 & 0.2040 \\ \hline
\end{tabular}
\caption{Model comparison and assessment metrics for the six models. CRPS is averaged over 26 weeks.}
\label{tab:loo_ic}
\end{table}

\begin{figure}
    \centering
    \includegraphics[width=1\linewidth]{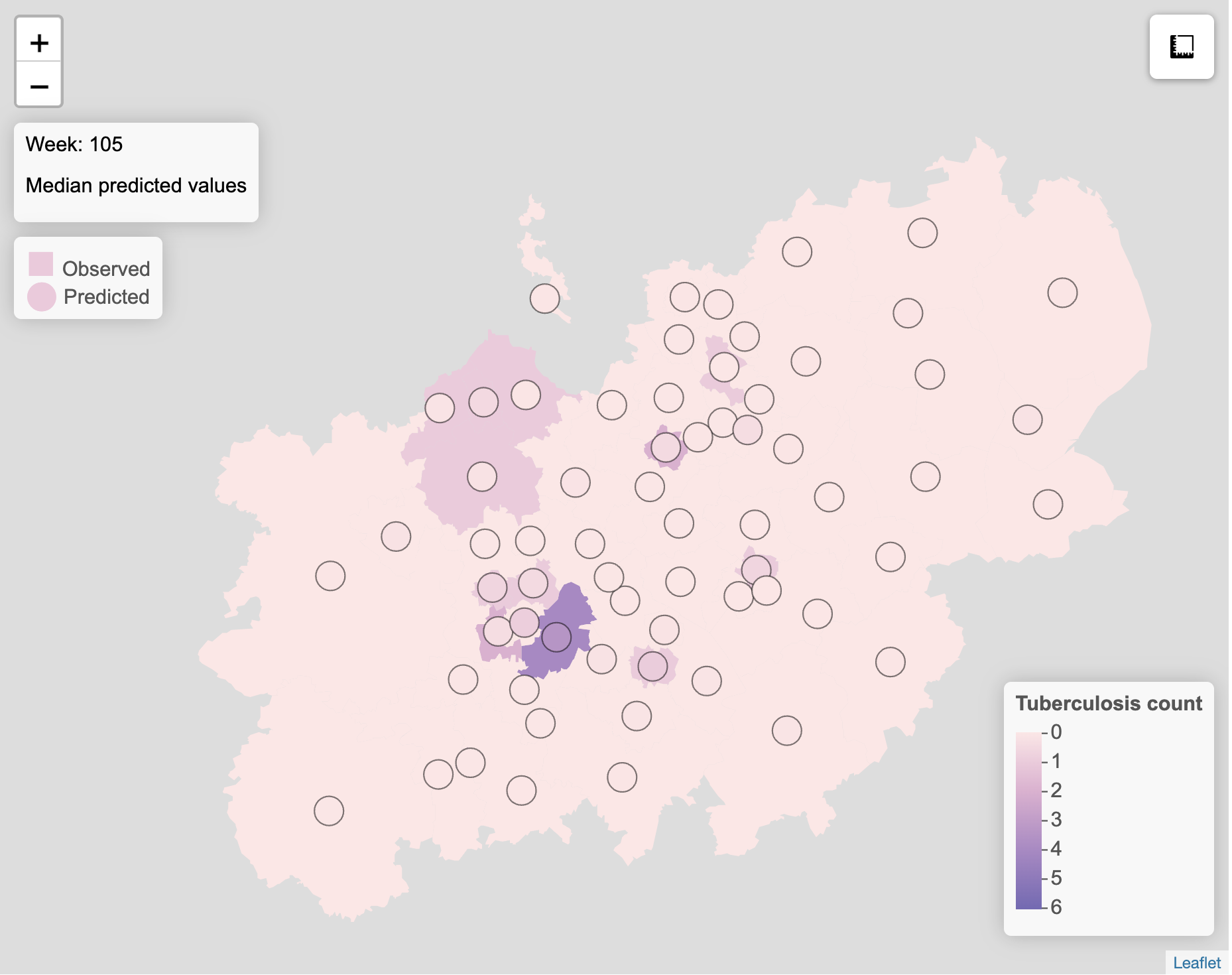}
    \caption{Map showing observed (shaded map) and median posterior predictions (shaded circles) of Tuberculosis cases for the East and West Midlands for the first week of 2024 using our best model (model 2). This map was made using Leaflet \citep{leaflet} with Shiny \citep{shiny}.}
    \label{fig:week105_model2}
\end{figure}

\section{Conclusions}
\label{sec:concs}

Gaussian processes are highly flexible nonprametric models that can be used for a variety of tasks in infectious disease modelling. Their flexibility comes in part from the analytic tractability of the models for both inverse modelling and prediction, but their use is dependent on inversion of potentially large design matrices.

Much previous work has been done to introduce computationally efficient and accurate approximations to these inversions. For example Vecchia approximation \citep{vecchia}; H-matrices \citep{geoga2020scalable,moran22} amongst others. As well as efficient linear algebra, the use of efficient computational hardware to speed up model fitting and prediction can drastically improve the practical fitting of GPs, and software enabling this with minimal user input and expertise is important.

Here we apply the GP models to a purely data-driven application to help explain and predict spatial and temporal variation in Tuberculosis across the East and West Midlands, England. We show that we can obtain accurate and consistent predictions of counts of Tuberculosis cases up to 26 weeks ahead, conducting the inference in a much more realistic timeframe than using existing CPU approaches, whilst estimating predictive uncertainty to help inform decision making.

The approach is very flexible to a wide variety of response distributions, with the kernels explaining the inherent correlation between monitoring locations (here PHE authority areas) and across time points. The separation of time and space also readily supports fitting across larger spatial and temporal regions. Additional correlation structures can be incorporated where required through addition or multiplication of kernels, but in this case study, relatively simplistic spatial Mat\`{e}rn kernels were sufficient to account for spatial variability. The use of the negative binomial distribution also accounts for additional variation beyond a Poisson model, which assumes a constant mean-variance relationship. We combined population density into the model through a background population effect. This could be further extended to account for heterogeneities in the population by incorporating additional demographic or environmental effects, either through the background offset or through the bias term of the GP.

{The approach was capable of predicting at least 26 weeks ahead without significant loss of predictive ability, suggesting the training data are sufficient to learn signal through space and time. The support for a periodic signal of tuberculosis spread in the East and West Midlands suggests seasonal variation in incidence that is relatively consistent across years, supporting previous work on TB globally \cite{sasonaltb}.}

In conclusion, the \texttt{greta.gp} package enables flexible hierarchical modelling of spatio-temporally indexed data on infectious disease incidence, and the combination with GPU hardware allows fitting and prediction over larger regions in realistic timeframes. This manuscript acts as a tutorial for the fitting of these flexible GPs in \texttt{R}, enabling practitioners to easily apply and tailor these to their own problems of interest. 




\bibliographystyle{elsarticle-num-names} 
\bibliography{main.bib}





\end{document}